\documentclass[11pt,a4paper]{article}
\pdfoutput=1
\usepackage{jinstpub}
\usepackage{lineno}

\usepackage{float}
\usepackage{gensymb}
\usepackage{url}
\usepackage{svg}
\usepackage{caption}
\usepackage{subcaption}
\usepackage{amsmath} 
\usepackage{array}
\usepackage{tabularx}
\usepackage{xcolor} 

\newcommand{\mum}{~\rm \upmu m}

\title{\boldmath In beam performances of the MIMOSIS-2.1 CMOS Monolithic Active Pixel Sensor}
\author[a,b,1]{M.~Deveaux\note{Corresponding author.}}
\author[c]{, Ali-Murteza Altingun}
\author[a]{, Julio Andary}
\author[a]{, Benedict Arnoldi-Meadows}
\author[c]{, Jerome Baudot}
\author[c]{, Gregory Bertolone}
\author[c]{, Auguste Besson}
\author[a]{, Norbert Bialas}
\author[a]{, Christopher Braun}
\author[c]{, Roma Bugiel}
\author[c]{, Gilles Claus}
\author[c]{, Claude Colledani}
\author[a,c]{, Hasan Darwish}
\author[c]{, Andrei Dorokhov}
\author[c]{, Guy Dozi\`ere}
\author[c]{, Ziad El Bitar}
\author[a,b]{, Ingo Fröhlich}
\author[c]{, Mathieu Goffe}
\author[a]{, Benedikt Gutsche}
\author[c]{, Abdelkader Himmi}
\author[c]{, Christine Hu-Guo}
\author[c]{, Kimmo Jaaskelainen}
\author[f]{, Oliver Keller}
\author[a]{, Michal Koziel}
\author[a]{, Franz Matejcek}
\author[a]{, Jan Michel}
\author[c]{, Frederic Morel}
\author[a]{, Christian M\"untz}
\author[c]{, Hung Pham}
\author[b]{, Christian Joachim Schmidt}
\author[a]{, Stefan Schreiber}
\author[c]{, Matthieu Specht}
\author[a,b,d]{, Joachim Stroth}
\author[a]{, Eva-dhidho Taka}
\author[c]{, Isabelle Valin}
\author[a]{, Roland Weirich}
\author[c]{, Y\"ue Zhao}
\author[e]{and Marc Winter}
\affiliation[a]{Institut für Kernphysik, Goethe-Universität Frankfurt, Max-von-Laue-Straße 1, 60438 Frankfurt (Germany)}
\affiliation[b]{GSI Helmholtzzentrum für Schwerionenforschung GmbH, Planckstraße 1, 64291 Darmstadt (Germany)}
\affiliation[c]{Universit\'e de Strasbourg, CNRS, IPHC UMR 7178, 67037 Strasbourg (France)}

\affiliation[d]{Helmholtz Forschungsakademie Hessen für FAIR, Max-von-Laue-Straße 12, 60438 Frankfurt (Germany)}
\affiliation[e]{Universit\'e Paris-Saclay, CNRS/IN2P3, IJCLab, 91405 Orsay (France)}
\affiliation[f]{Facility for Antiproton and Ion Research GmbH, Planckstraße 1, 64291 Darmstadt (Germany)}

\emailAdd{m.deveaux@gsi.de}
\abstract{MIMOSIS is a CMOS Monolithic Active Pixel Sensor developed to equip the  Micro Vertex Detector of the Compressed Baryonic Matter (CBM) experiment at FAIR/GSI. The sensor will combine an excellent spatial precision of 5 \textmu m with a time resolution of 5 \textmu s and provide a peak hit rate capability of $\mathrm{\sim 80~ MHz/cm^2}$. To fulfill its task, MIMOSIS will have to withstand ionising radiation doses of $\sim 5~ \mathrm{MRad}$ and fluences of $\sim 7 \times 10^{13}~\mathrm{n_{eq}/cm^2}$ per year of operation.  \\
This paper introduces the reticle size full feature sensor prototype MIMOSIS-2.1, which was improved with respect to earlier prototypes by adding on-chip grouping circuts and by improving the analog power grid. Moreover, it features for a first time a $50 \mum$ epitaxial layer, which is found to improve the performances of the non-irradiated device significantly. We discuss the in beam sensor performances as measured during beam tests at the CERN-SPS.}

\begin{document}
\maketitle
\flushbottom

\section{Introduction}
\label{sec:intro}
The MIMOSIS CMOS Monolithic Active Pixel Sensor (CPS) will equip the Micro Vertex Detector (MVD) \cite{MVD} of the Compressed Baryonic Matter (CBM) experiment \cite{CBM}, which is being built at the FAIR facility in Darmstadt, Germany. It must combine a spatial precision of \textasciitilde 5 µm with a thickness of \textasciitilde 50 µm to reach the targeted material budget of 0.3-0.5\%~$X_{0}$ per detector station. Moreover, a mean rate capability of slightly below 20 MHz/cm$^2$ in average and of up to 80 MHz/cm$^2$ for few \mbox{10 µs} and a to 5 MRad and $7\times 10^{13}~\mathrm{n_{eq}/cm^2}$ per CBM year of operation is needed. 

As introduced in \cite{Paper:MIMOSIS-0, MIMOSISHasan}, MIMOSIS relies on an industrial 180 nm CMOS imaging process. It will feature 1024 columns of 504 pixels, the latter being $30  ~\mathrm{\upmu m} \times 27 ~\mathrm{\upmu m}$ wide. Each pixel integrates a complete amplifier / shaper / discriminator chain, inspired by the read-out architecture of the ALPIDE sensor equipping the ALICE-ITS2 detector \cite{ALPIDE}. If the signal charge related to a particle hit exceeds the discriminator threshold, the analogue front-end samples and holds this information. At the end of a 5 µs long integration time (frame), the hit information is forwarded to an output memory of the pixel. The output memories of the 1008 pixels composing two neighboring columns are read out by a common priority encoder, which performs a zero-suppression. The readout-priority of the pixels in the double-columns follows a meandering pattern. Typically more than one pixel fires per impinging particle. Once a firing pixel is recognized, a specific on-chip pixel grouping circuit scans the three consecutive pixels in the sense of readout priority. The information on these four pixels are encoded into one common 16-bit word. Possible further pixels of a cluster are processed separately. The data is sent out via two layers of data concentration buffers and a central elastic buffer through up to eight 320 Mbps differential data links. 

\begin{figure}[h]
\centering
\begin{minipage} [c]{4.8cm}
\includegraphics[width=4.8cm]{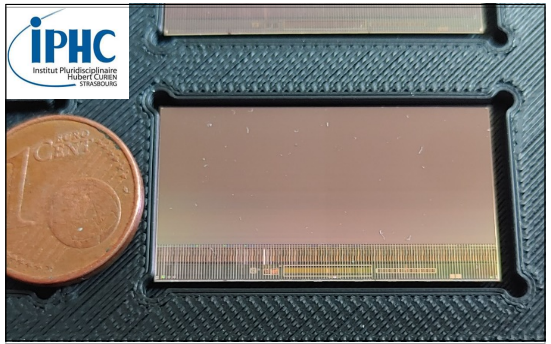}
\vspace{.55cm}
\caption{Photograph of the MIMOSIS-1 sensor.}
\label{fig:MIMOSIS-1}
\end{minipage}
\hspace{0.5cm}
\begin{minipage}[c]{9.2cm}
\centering
\includegraphics[width=9.2cm]{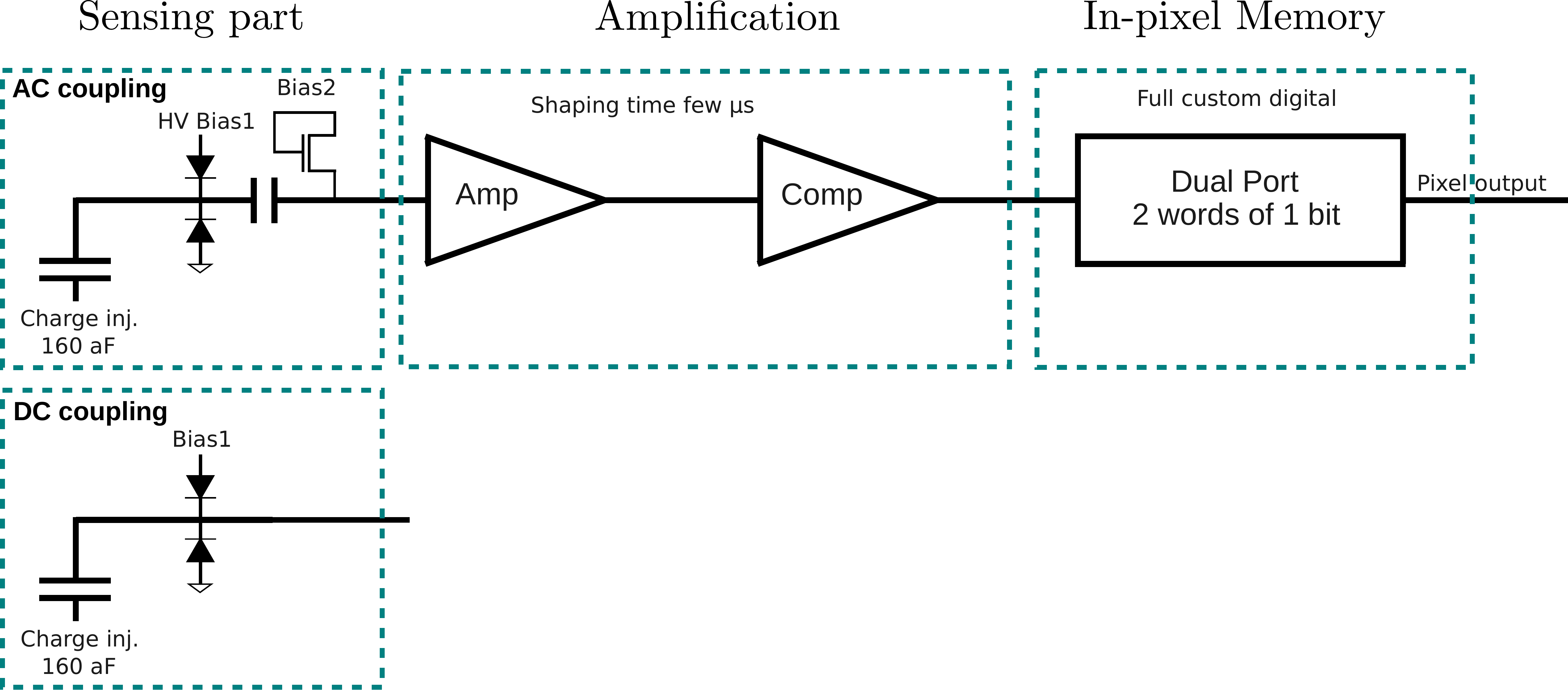}
\caption{Block diagram of the AC (top) and DC (bottom) coupled pixels. }
\label{fig:pixel_layout}
\end{minipage}
\end{figure}

MIMOSIS-2 forms a full feature prototype of MIMOSIS. Appearing like the successful full size prototype MIMOSIS-1 (see Fig.~\ref{fig:MIMOSIS-1}), it hosts new features like the above mentioned pixel grouping, clock triplication as needed for improved SEE tolerance, an improved single bit flip correction fixing the issues reported in \cite{PaperBen} as much as an improved analogue power net intended to further reduce the fixed pattern noise. It hosts 128 columns of DC-coupled pixels as used in ALPIDE, and three matrices with 384, 384 and 128 columns of AC-coupled pixels (see Fig. \ref{fig:pixel_layout} for details). All pixels of a sub-matrix share their steering voltages/currents and thus feature a common threshold. The size of the sub-matrices was relevant for MIMOSIS-1 only, but kept for the sake of risk minimization. 

\begin{figure}[t]
     \centering
          \begin{minipage}{0.33\textwidth}
         \centering
         \hspace{5mm}
         \includegraphics[width=\linewidth,height=3.4cm]{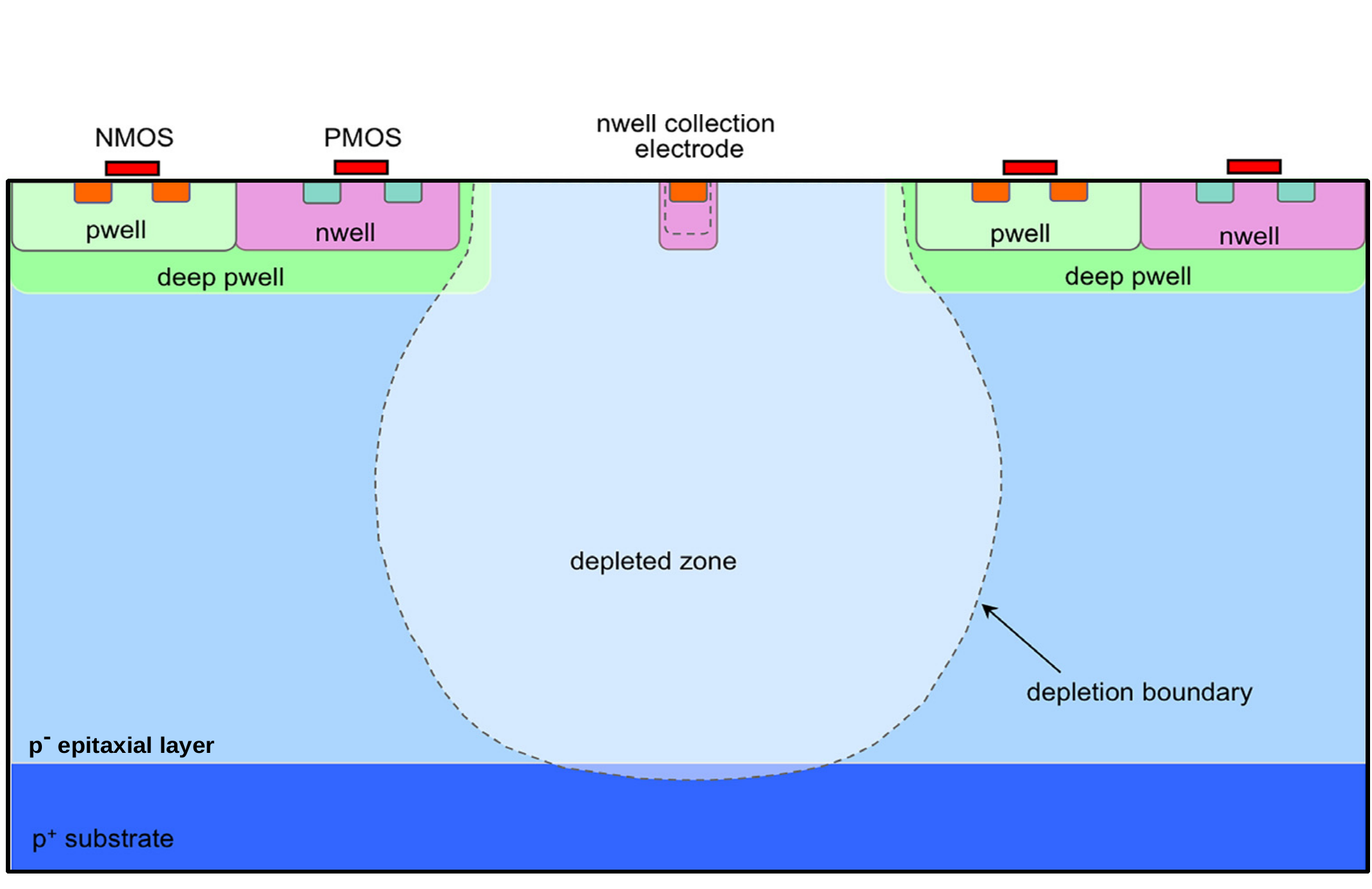}
      \end{minipage}
     \begin{minipage}{0.33\textwidth}
         \centering
         \includegraphics[width=\linewidth]{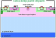}
     \end{minipage}
     \hfill
     \begin{minipage}{0.3\textwidth}
     \hspace{0.8cm}
         \caption{Simplified cross-section of the standard (left) and p-stop (right) sensing node options used in MIMOSIS-2.1. From \cite{walter_paper,doublemodified_paper}.}
        \label{fig:3splits}
    \end{minipage}
\end{figure}

MIMOSIS-2.1, the error corrected version of MIMOSIS-2, was manufactured with different epitaxial layer profiles: the partially depleted standard and the likely fully depleted p-stop layout \cite{walter_paper, doublemodified_paper} (see Fig. \ref{fig:3splits}) were each combined with wafers with 25 and 50 µm thick epitaxial layer. The n-gap pixel available on MIMOSIS-1 had shown no added value w.r.t the p-stop pixel and was therefore discontinued. All results shown refer to MIMOSIS-2.1, also denoted as Mi2 or MiSIS-2.1 in some captions/plots.

\section{Estimate of the pixel gain}

Expressing noise and thresholds of the pixel in units of electrons requires calibration. This was done by injecting a voltage pulse $\Delta U$ to an in-pixel charge injection capacitor (see Fig.~\ref{fig:pixel_layout}), which converts it to a signal equivalent charge according to $\Delta Q = C_{\mathrm{inj}} \cdot \Delta U$. $C_{\mathrm{inj}}$ was tuned to $160~\mathrm{aF}=1~\mathrm{e/mV}$ by layout extraction but the true value remained uncertain due to complex layout and the tiny absolute value. Still, in lack of better knowledge, we used this value in our previous works. 

$ C_{\mathrm{inj}}$ was measured with the test structure CE18. CE18 hosts MIMOSIS-1/2-like pixels with identical front end, collection node and injection circuit and additional direct access to both, the amplifier and comparator outputs; and SB-pixels with source follower as described e.g.~in \cite{MichaelHabil}. The latter allowed for comparing the response to $\Delta Q=1640~e$ as injected with the $5.9~\mathrm{keV}$ X-rays of a $^{55}$Fe-source (see \cite{MichaelHabil} for details) with the one to the charge injection system. The MIMOSIS-1/2 amplifiers saturate at this $\Delta Q$ but their ToT could be calibrated and exploited. 

The results based on pixels with 25 µm epitaxial layer indicate $C_{\mathrm{inj}}$ to deviate substantially from its design value: A $C_{\mathrm{inj}}$ of $1,86~\mathrm{e/mV}$ and $2.02~\mathrm{e/mV}$ is observed for the DC and AC pixels, respectively, with no significant differences between standard and p-stop pixels. The individual gain of $>90\%$ of the pixels remains within $\pm 10\%$ around the reported mean values. The new gain factors should be used to re-calibrate the electron scales of our previously published papers. They may also partially explain the correction required for matching data and simulations of the very similar ALPIDE sensor, which required a factor of 1.63 according to \cite{SuljicPhD}.

\section{The MIMOSIS-2.1 beam telescope and data analysis}
The detection efficiency and spatial resolution of MIMOSIS-2.1 was measured in the $\sim  120~\mathrm{GeV}/c$ pion beam of the CERN-SPS. The reference planes of our MIMOSIS-1-telescope \cite{MIMOSISHasan} were replaced by 300 µm thick  MIMOSIS-2.1 sensors with standard pixels and 25 µm epitaxial layer. Due to a failing cooling system, we operated at non-controlled room temperature (likely \mbox{$T_{sensor}\approx 30-40^{\circ}\mathrm{C}$)}. The data was processed with the TAF data analysis package \cite{TAF} as explained in \cite{MIMOSISHasan}.
The resolution was derived from the distance between the interpolated beam particle hit position and the closest hit found in the DUT within a 100 µm search window. 
The center-of-gravity of the hit clusters was computed to improve the spatial resolution of all detector planes.

\begin{figure}[t]
\centering
  \begin{minipage}{0.6\textwidth}
   \includegraphics[width=\columnwidth, trim={50 20 90 60}, clip]{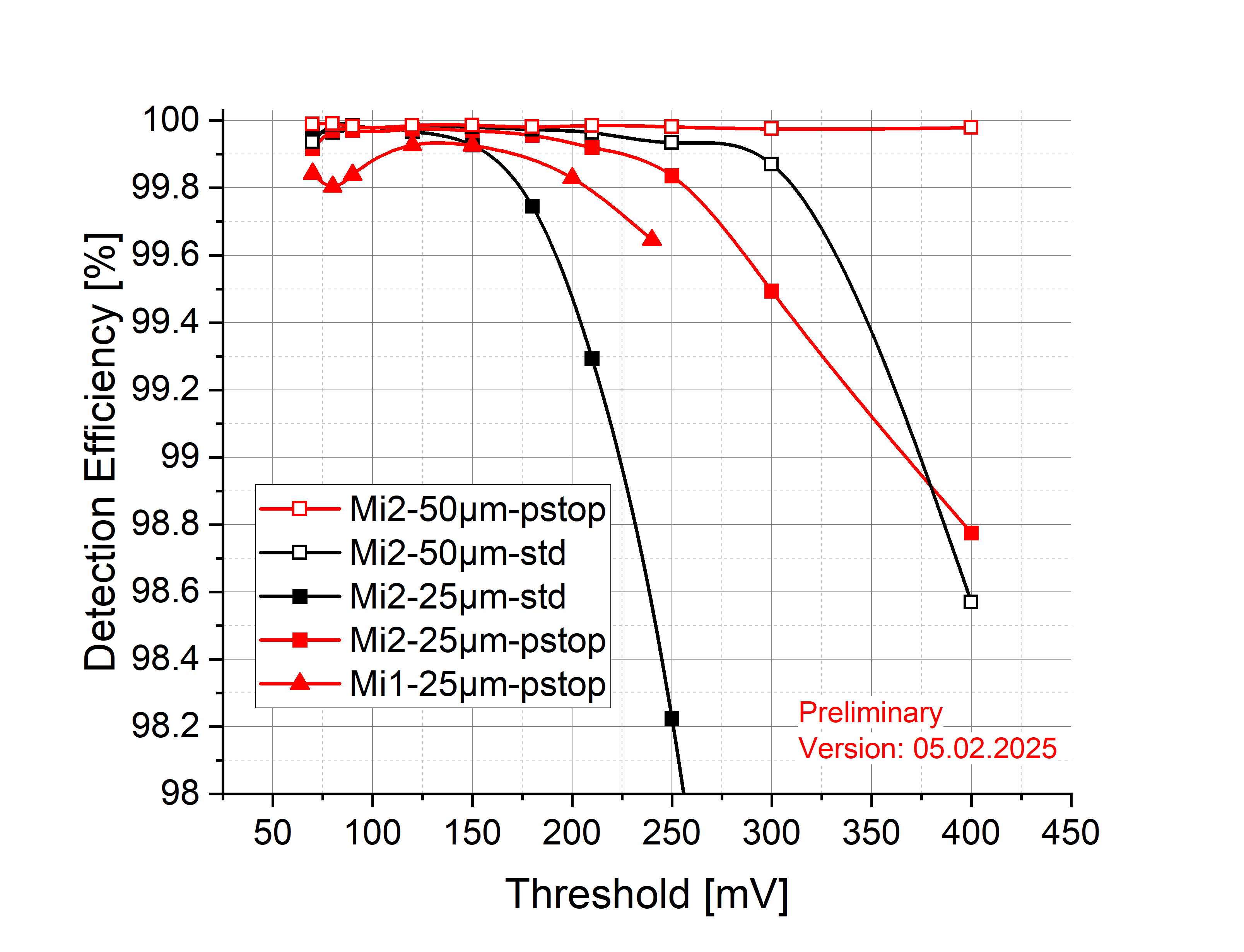}
  \end{minipage} 
  \hfill
  \begin{minipage}{0.35\textwidth}
   \caption{Detection efficiency of the pixels of MIMOSIS-2.1 as compared to the p-stop pixel of MIMOSIS-1. Lines to guide the eye. Note the color/symbol code:
    \label{fig:EffVsThreshold}  }

    \begin{tabular}{|l|l|}
          \hline
           standard pixel & \textcolor{red} {p-stop pixel} \\
           \hline
            $\blacksquare \blacktriangle$ 25 µm epi &  $\Box  \triangle$ 50 µm epi \\
           \hline
            $\triangle \blacktriangle$ MiSIS-1&$\Box\blacksquare $ MiSIS-2.1 \\
            \hline
       \end{tabular}

  \end{minipage}

\end{figure}

The telescope-precision was estimated based on the software "The Telescope Optimizer" \cite{telescopeOptimizer}. As the spatial precision of the reference planes was initially uncertain, a DUT identical to the sensors of the reference planes was used. The indicated residuals were used as precision of the reference planes. Hereafter, the indicated track precision of the telescope was folded out of the residuals of the DUT and the updated, indicated sensor precision was considered for the telescope. This approach converged after few iterations. After converging, the telescope precision obtained was fixed and used for estimating the properties of the consecutive DUTs.


\section{Beam test results}
\label{sec:beamtestresults}

The beam tests focused on AC pixels, which were hosted by 300 µm thick sensors and operated with a top bias (HV) of 10V and a back bias of $-$1V.  The results are shown in Fig. \ref{fig:EffVsThreshold} and compared with results from the p-stop pixel of a 60 µm thick MIMOSIS-1 with 25 µm thick epitaxial layer. All pixels to provide a detection efficiency of $>99.9\%$. MIMOSIS-2.1 exceeds the already impressive efficiency of MIMOSIS-1. This progress is currently attributed to the enhanced biasing network of MIMOSIS-2.1, which is expected to reduce the fixed pattern noise. However, this has yet to be confirmed. Best detection efficiency is seen for the p-stop pixels with 50 µm epitaxial layer. The observed dark rates without masking remain below the $\sim 10^{-10}/$pixel detection limit in most cases and at least one order of magnitude below the required $\sim 10^{-5}/$pixel even at lowest thresholds.

\begin{figure}[tbp]
     \centering
     \begin{subfigure}[b]{0.49\textwidth}
         \centering
         \includegraphics[width=\textwidth,trim={60 20 80 40}, clip]{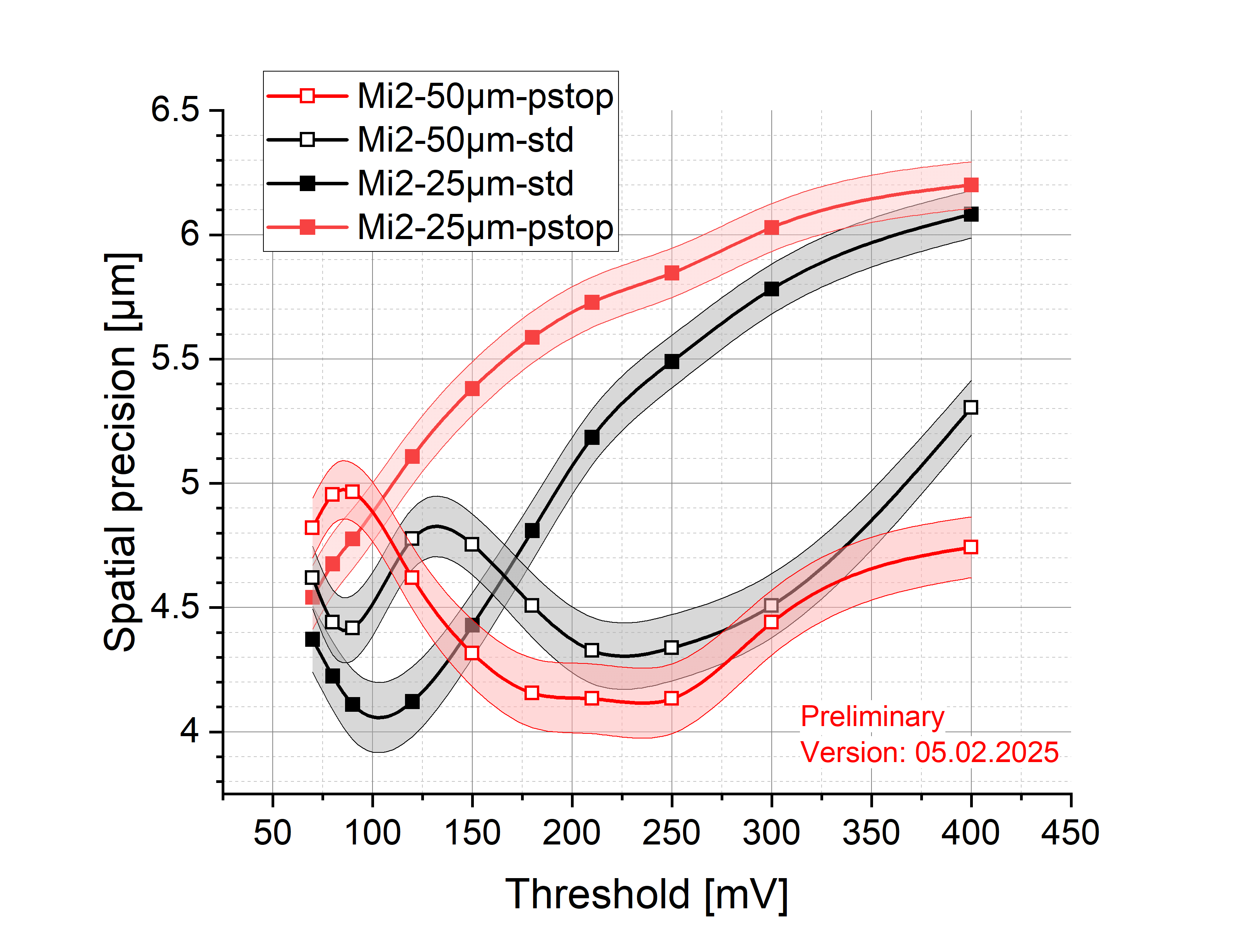}
         \caption{Spatial precision vs.~threshold.}
         \label{fig:Resolution}
     \end{subfigure}
     \hfill
     \begin{subfigure}[b]{0.49\textwidth}
         \centering
         \includegraphics[width=\textwidth,trim={60 20 80 60}, clip]{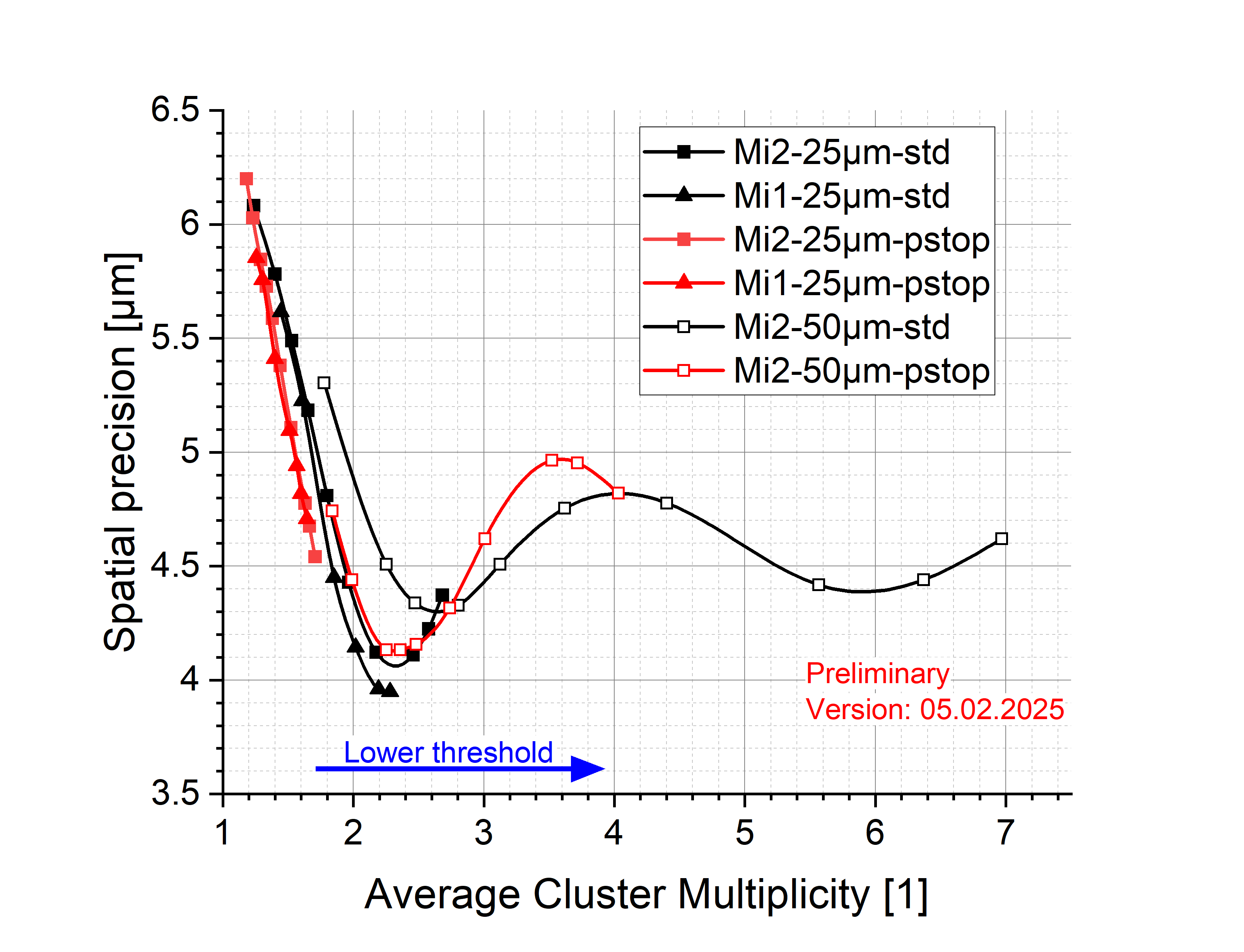}
         \caption{Spatial precision vs.~cluster multiplicity.}
         \label{fig:ResVsMultiplicity}
     \end{subfigure}
     
     \begin{minipage}{0.49\textwidth}
       \caption{Spatial precision of different MIMOSIS pixels as function of threshold (a) and pixel multiplicity (b) for the shorter side of the $27\times 30 ~\mathrm{\upmu m^2}$ pixel. A $\sim 10\%$ worse precision is seen for the other dimension (not shown). Lines to guide the eye.}\label{fig:ResolutionPlots}  
     \end{minipage}
     \hfill
     \begin{minipage}{0.45\textwidth}
       \begin{tabular}{|l|l|}
          \hline
           standard pixel & \textcolor{red} {p-stop pixel} \\
           \hline
            $\blacksquare \blacktriangle$ 25 µm epi &  $\Box  \triangle$ 50 µm epi \\
           \hline
            $\triangle \blacktriangle$ MiSIS-1&$\Box\blacksquare $ MiSIS-2.1 \\
            \hline
       \end{tabular}
       \end{minipage}
\end{figure}

The spatial precision is displayed for the different pixel designs as function of the threshold in Fig.~\ref{fig:Resolution}. All pixels except the p-stop pixel with 25 µm epitaxial layer exhibit unexpected fluctuations in a band between 4 and 5 µm. We consider that this pattern is a feature of the center-of-charge calculations, which may be more or less efficient depending on the average cluster multiplicity. This hypothesis was checked by plotting the spatial precision of the pixels as function of the measured multiplicity (see \mbox{Fig. \ref{fig:ResVsMultiplicity}).} This plot was complemented by previously recorded data from MIMOSIS-1. We observe that, independently of the pixel details, the spatial resolution is first improved with increasing cluster multiplicity ($C_M$). This known trend stops at an optimum located around $C_M\approx2.4$, which is about the maximum cluster size reported for most pixels of MIMOSIS and similar sensors. However, the (anticipated) high S/N of some MIMOSIS-2.1 pixels allows for tuning them to higher $C_M$, which reveals a local maximum at around $C_M\approx 3.8$ and a second, less pronounced optimum at $C_M \approx6$. Remarkably, all pixels flavors follow this pattern despite they were measured with individual alignment. This suggests a plausibly geometrical mechanism, which is independent of the detailed pixel type.  Its nature remains to be studied.

\section{Summary and conclusion}

MIMOSIS-2.1, the first full scale full feature prototype of the future MIMOSIS CPS was tested in beam. The so far most promising pixel (p-stop, 25 µm) of MIMOSIS-2.1 is found to reproduce and even exceed the excellent detection efficiency results obtained from the previous MIMOSIS-1 prototype. Combining this sensing element with a novel, 50 µm thick epitaxial layer mostly exploits the little remaining room for improvement in terms of detection efficiency. Moreover, we observe an increase in the cluster multiplicity, which creates a spatial precision matching the targeted \mbox{5 µm} for suited threshold settings. A systematic pattern in the relation between spatial precision and cluster multiplicities of the pixels is observed and remains to be studied. 

The absolute calibration of the threshold values in units of electrons was reviewed exploiting dedicated test structures. We find that the gain of the underlying charge injection system diverges by a factor of about two from its design value. Threshold and noise values as shown in earlier communications should be evaluated with this factor in mind. This changes the numerical values shown therein but does not compromise or modify the conclusions of those works.

No claim on the radiation tolerance of the new sensing element is made. In case the fully depleted 50 µm thick p-stop pixel keeps its benefits after being exposed to the targeted dose of $10^{14}~\mathrm{n_{eq}/cm^2}$, this pixel might come out as the option of choice for the CBM-MVD. 

\acknowledgments
 This work has been funded and supported by the German federal Ministry of Education and Research (BMBF), the European network for developing new horizons for RIs (Eurizon), HFHF and HGS-HIRe for FAIR.


\end{document}